\documentclass[preprint,12pt]{elsarticle}

\usepackage{amsmath,amssymb,amsfonts}
\usepackage{comment}

\usepackage{subcaption}
\usepackage[margin=1in]{geometry}

\usepackage{algorithmic}
\usepackage{graphicx}
\usepackage{algorithm,algorithmic}
\usepackage{svg}
\usepackage{listings}
\usepackage{xcolor}
\usepackage{inconsolata} %
\usepackage{url}
\usepackage{longtable}
\usepackage{multirow}
\usepackage{tikz}
\usetikzlibrary{positioning}
\usepackage{lineno}
\usepackage{pifont}
\newcommand{\xmark}{\ding{55}} %

\definecolor{myred}{rgb}{0.8, 0.0, 0.0}
\renewcommand{\arraystretch}{1.4}

\journal{Information Sciences}

\begin{document}

\begin{frontmatter}

\title{Shill Bidding Prevention in Decentralized Auctions Using Smart Contracts}

\author[1,2]{M.A. Bouaicha}
\author[2]{G. Destefanis}
\author[1]{T. Montanaro}
\author[3]{N. Lasla}
\author[1]{L. Patrono}

\affiliation[1]{organization={University of Salento}, 
            city={Lecce}, 
            country={Italy}}

\affiliation[2]{organization={Brunel University of London}, 
            city={London}, 
            country={United Kingdom}}

\affiliation[3]{organization={National School of Artificial Intelligence (ENSIA)}, 
            city={Algiers}, 
            country={Algeria}}

\begin{abstract}
In online auctions,  fraudulent behaviors such as shill bidding pose significant risks. This paper presents a conceptual framework that applies dynamic, behavior-based penalties to deter auction fraud using blockchain smart contracts. Unlike traditional post-auction detection methods, this approach prevents manipulation in real-time by introducing an economic disincentive system where penalty severity scales with suspicious bidding patterns. The framework employs the proposed Bid Shill Score (BSS) to evaluate nine distinct bidding behaviors, dynamically adjusting the penalty fees to make fraudulent activity financially unaffordable while providing fair competition.
 
The system is implemented within a decentralized English auction on the Ethereum blockchain, demonstrating how smart contracts enforce transparent auction rules without trusted intermediaries. Simulations confirm the effectiveness of the proposed model: the dynamic penalty mechanism reduces the profitability of shill bidding while keeping penalties low for honest bidders. Performance evaluation shows that the system introduces only moderate gas and latency overhead, keeping transaction costs and response times within practical bounds for real-world use. The approach provides a practical method for behaviour-based fraud prevention in decentralised systems where trust cannot be assumed.

\end{abstract}


\end{frontmatter}

\section{Introduction}
Online auctions have become a widely used service, allowing buyers and sellers to trade goods remotely through platforms like eBay and Yahoo {\cite{MAJADI20171}. Their accessibility has contributed to significant growth, reflected in sales figures from major auction houses. According to ArtTactic~\cite{barrons2022auction}, online auction sales reached \$1.35 billion in 2021, a 28.2\% increase from 2020, with projections indicating further growth of \$1.9 billion by 2026 \cite{Ye202313909}.

Despite their wide use, centralized auction systems are still susceptible to fraudulent behavior, affecting sellers, bidders, and auctioneers too. One issue is Bid Shielding, where a set of bidders collude to get the item at a lower price by placing low bids~\cite{MAJADI20171}. Another significant fraud is Shill Bidding (SB), the focus of this work, where sellers or their associates place fake bids to inflate final prices~\cite{IngebretsenCarlson2022341}.
These fraudulent activities damage platform trust and user engagement.

Current SB detection and prevention systems struggle to between legitimate and fraudulent bidding activities~\cite{MAJADI20171}. The patterns used to identify shill bidders are not always reliable, and there is no definitive way to determine SB with certainty~\cite{MAJADI20171}. The ability to create multiple pseudonymous accounts makes detection even harder, making collusion and repeated manipulation possible~\cite{wang2004shill}. These challenges reflect the need for an auction mechanism that discourages SB rather than relying on reactive approaches~\cite{wang2004shill}.

Blockchain has been proposed as a solution to enhance fairness and transparency on online auctions~\cite{DAI2024100201}. Its transparency helps resolve the issue of information asymmetry faced in classic central auctions, where part of the bidders have access to more information than other~\cite{bloomenthal2021asymmetric}.
This limits the scope for collusion and improves the identification of fraudulent activities such as SB~\cite{DAI2024100201}.
Furthermore, blockchain provides timestamping to record transaction time and smart contracts~\cite{Xiong2021}, which are secure, and self-executing code that automates agreements without relying on a trusted third party~\cite{bashir2020mastering}. Researchers have explored blockchain-based auction systems to support decentralization and automation~\cite{wu2018cream}, with applications including tokenizing auctioned items and real-time bidding~\cite{liu2024blockchain}. However, despite these developments, fraud prevention in blockchain-based auctions remains underexplored, and existing solutions have failed to address these threats~\cite{shi2022integration}.

This paper presents a blockchain-based English auction framework that employs smart contracts to discourage SB through a dynamic penalty mechanism. The framework includes a penalty model that tracks nine different bidding patterns continuously and charges penalty fees depending on bidding actions.
The following research questions guided the design and implementation of the framework:
\begin{enumerate}
    \item What are the specific SB behaviors that can be identified and mitigated through smart contracts in blockchain-based auctions while maintaining scalability and cost-efficiency?
    \item How can we design a dynamic transaction fee mechanism within blockchain-based auction systems to discourage SB while ensuring that honest bidders are not subjected to excessive fees?
\end{enumerate}
Building on these questions, this paper highlights the lack of prior research comparing bidding patterns in decentralized blockchain-based auctions and traditional centralized auctions. To bridge this gap, we first need a system that can monitor these patterns within the context of blockchain-based auction systems. Based on this, we design and implement a framework in which the smart contract acts as a consensus mechanism between bidders and sellers. The framework applies predefined rules to deter fraudulent bidding through a dynamic bid fee mechanism inspired by transaction fee models in blockchain networks. This mechanism adjusts penalties in real time based on a bidder's activity.
The main contributions provided in this paper are:

\begin{enumerate}
\item A blockchain-based English auction framework with an integrated anti-shill bidding mechanism.
\item A fraud prevention system based on two models: the Bid Shill Score (BSS) for evaluating bidder behavior and a dynamic penalty fee model that increases transaction costs for suspected fraudulent bidders.
\end{enumerate}

To achieve these contributions, we first identified common SB patterns as reported in the existing literature. We then measured each of these patterns with an associated metric. Based on these metrics, we developed a dynamic penalty system that gets updated in real time according to the BSS. This mechanism was implemented within a smart contract in order to allow decentralized enforcement of these rules.  Finally, we discussed the theoretical advantages and trade-offs of using smart contracts for on-chain prevention of SB. This approach aims to fill existing research gaps and introduces a cost-effective, scalable solution for SB prevention in decentralized online auctions.

The remainder of this paper is organized as follows. Section 2 defines important concepts for the proposed work, such as auction types and SB fraud. Section 3 reviews existing work on SB detection and blockchain-based auctions. Section 4 details the proposed framework, covering the smart contract design, penalty the proposed framework, smart contract design, penalty mechanism, and bidding patterns metrics. Section 5 outlines the implementation process.
Section 6 discusses the framework's performance and scalability, highlighting the ability to identify advanced SB strategies while maintaining efficiency. Section 7 highlights the system limitations, and Section 8 concludes the paper with directions for future research.
\section{Background}
This section outlines the concepts relevant to the proposed work. First, it describes different types of auctions, their mechanisms, and common applications.  Then, we examine SB, its motivations, and its impact on auction integrity.

\subsection{Types of auctions}
An auction is a structured sales process in which potential buyers engage in competitive bidding for various goods or services~\cite{shi2022integration}. It usually involves four main elements: (1) a seller who owns the item, (2) one or more bidders who are interested to buy it, (3) the auctioned item, and (4) an auctioneer who conducts the procedure. Auctions are classified based on factors such as the bidding process, the number of items, and the visibility of participants and bids~\cite{10.1145/1883612.1883617}.

Based on bidding mechanisms, auctions can be open-outcry, in which bids are placed publicly, or sealed-bid, in which offers are placed privately~\cite{shi2022integration}. Both auctions have their characteristics and are suitable for different applications. The most commonly used auction models are described below and illustrated in Figure~\ref{fig:1}.

\begin{figure*}[htbp]
\centerline{\includegraphics[width=0.85\textwidth]{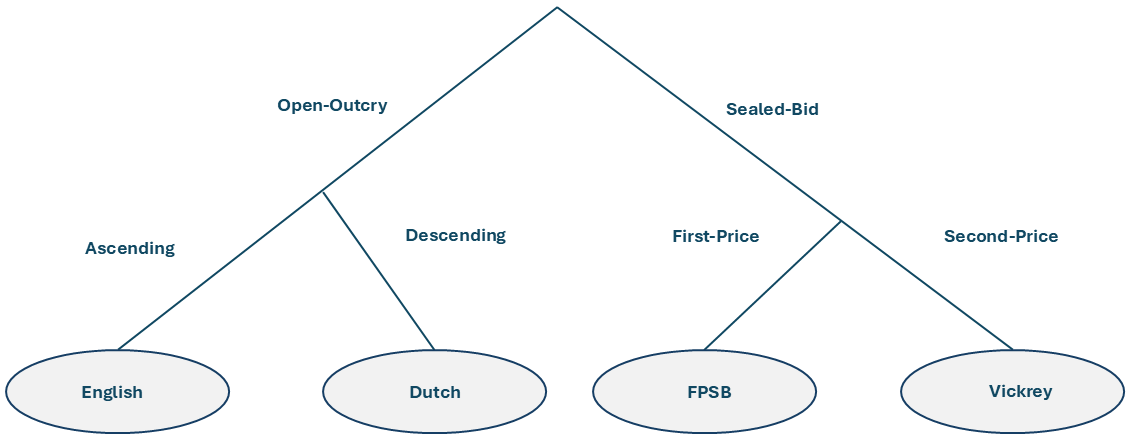}}
\caption{Summary of auction types}
\label{fig:1}
\end{figure*}

\begin{enumerate}
    \item \textbf{English Auction:} Also known as an open-outcry ascending-price auction~\cite{omar2021implementing}, the English auction, as shown in Figure~\ref{fig:eng-auc}, begins with a low starting price that increases as buyers place higher bids. The process continues until no further bids are received within the allotted time, and the highest bidder wins~\cite{10.1257/jep.3.3.3}. This auction type is highly transparent, as all participants can see the current highest bid~\cite{shi2022integration}.

    \item \textbf{Dutch Auction:} A Dutch auction, defined as an open-outcry descending-price auction, is where the auctioneer begins with a high price and decreases it progressively until the bid is accepted by someone. This auction is typically used for selling multiple identical items quickly
    ~\cite{omar2021implementing}.

    \item \textbf{First-Price Sealed-Bid Auction (FPSB):} In an FPSB, all bidders place their bids at the same time and in private, and each is allowed only one bid~\cite{omar2021implementing}. The auction ends with the determination of the highest bidder as the winner. This type requires very strategic planning because the bidders have to get the best out of both submitting a competitive bid and not overspending~\cite{10612625}.

    \item \textbf{Second-Price Sealed-Bid Auction (Vickrey Auction):} Similar to FPSB, sealed bids are placed by bidders simultaneously. However, in this format, the bidder with the highest bid wins but pays the price of the second-highest bid~\cite{MAJADI20171}. This approach encourages honest bidding because bidders have fewer incentives to overbid~\cite{10.1257/jep.14.3.183}.

\end{enumerate}
    \begin{figure*}[htbp]
    \centerline{\includegraphics[width=1\textwidth]{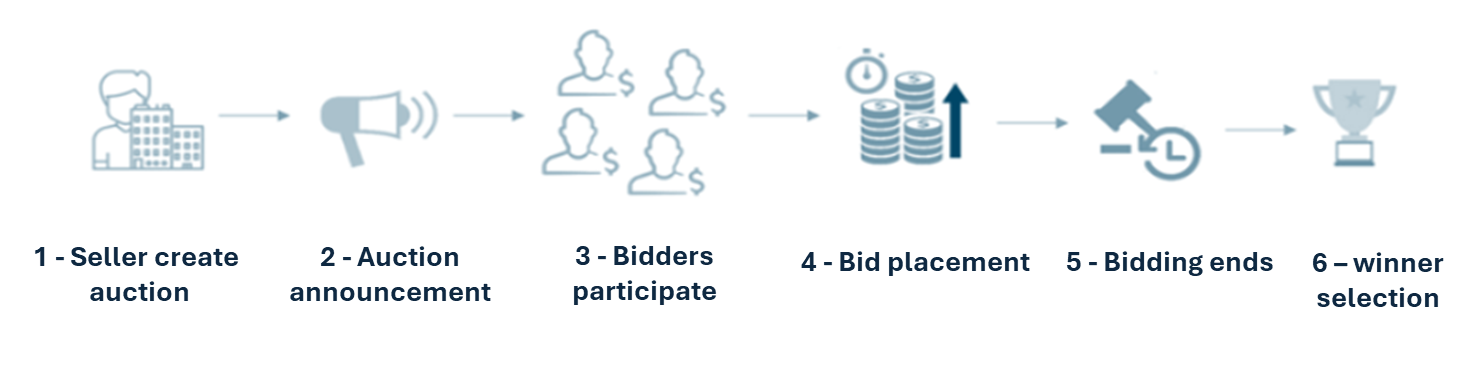}}
    \caption{English auction process}
    \label{fig:eng-auc}
    \end{figure*}

Each auction type has advantages and limitations, making it important to select the format that best suits the needs of buyers and sellers. However, all auctions are vulnerable to fraudulent practices that damages fairness and efficiency. SB is particularly common in English auctions, where open bidding makes it difficult to distinguish between fraudulent and legitimate participants. The similarities in bidding patterns makes detection challenging. The following section examines this fraudulent practice in more detail.

\subsection{Shill bidding}
Shill bidding, or ``shilling'', is the submission of false bids in order to inflate the final sale price. This manipulation is usually carried out by the seller in collusion with other parties or through several fake accounts~\cite{MAJADI20171}. Shill bidders aim to drive up the auction price, forcing legitimate bidders to place higher bids. This practice can be classified into two main types:

\begin{enumerate}
    \item \textbf{Competitive Shill Bidding:} In this strategy, shill bidders submit numerous small, incremental bids to falsely move the auction price higher, giving the impression of active competition. This strategy keeps up the pressure on legitimate bidders to raise their bids until the price reaches a targeted level. The shill bidder avoids winning the auction but succeeds in inflating the final sale price~\cite{MAJADI20171}.
    
    \begin{figure*}[htbp]
    \centerline{\includegraphics[width=1\textwidth]{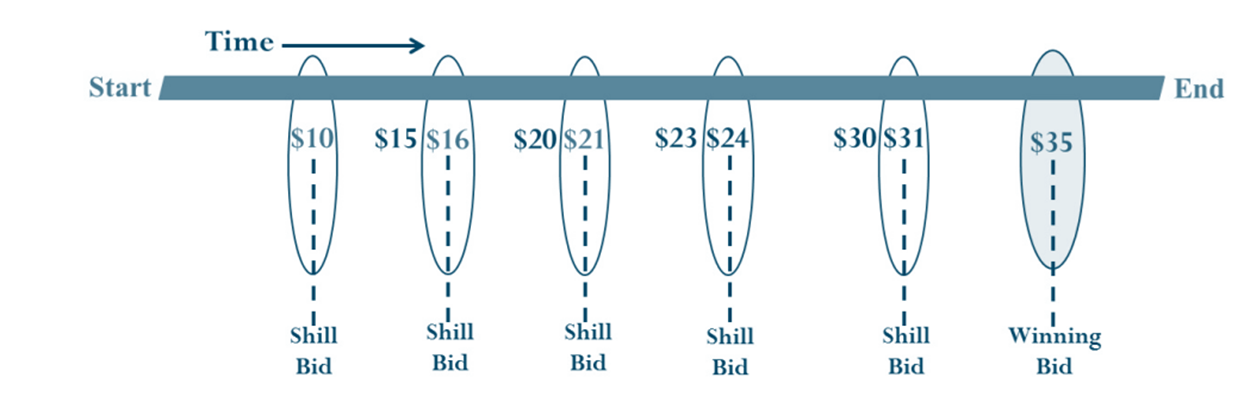}}
    \caption{Example of shill bidding scenario}
    \label{fig:shilling}
    \end{figure*}

    Figure~\ref{fig:shilling} illustrates an example where a shill bidder places multiple small bids immediately after a legitimate bidder. This forces the legitimate bidder to increase their bid during the auction, which causes paying more than necessary to win the auction.

    \item \textbf{Reserve Price Shilling:} This form of shilling targets auction platforms that charge listing fees based on the seller's declared reserve price. Sellers may list items with lower reserve prices to minimize fees, then use shill accounts to inflate the price to their actual desired value. If bidding remains too low, the seller may use shill accounts to ``buy back'' the item, avoiding selling at an unfavorable price~\cite{MAJADI20171}.

\end{enumerate}

Detecting SB is challenging due to the ease of creating anonymous accounts in online auctions, which makes it hard to distinguish between fraudulent and legitimate bids. Furthermore, online auction platforms usually do not disclose their monitoring methods, reducing transparency.

Given the impact of competitive SB, this study focuses on its prevention within decentralized auctions. Our proposed system utilizes smart contracts to create an open and tamper-evident auction environment and dynamically penalizes suspicious behavior according to predefined metrics. This approach aims to limit the effectiveness of competitive SB and improve the auction's integrity.

\section{Related Works}
While prior research has proposed various detection mechanisms, there has been a lack of studies on using smart contracts to prevent shill bidding and analyzing bidding patterns in decentralized auction systems. To bridge this gap, we first examine, SB detection and prevention strategies. Next, we review the blockchain applications in auctions, particularly its role in improving transparency and security, while identifying the open challenges in addressing auction fraud, especially in open-bid mechanisms. Finally, we outline the remaining gaps in current research that this work aims to address.

\subsection{Shill bidding detection and prevention mechanisms} 
Shill bidding has long been a challenging issue in online auctions. Research efforts have primarily focused on identifying and categorizing SB patterns, including early bidding, successive outbidding~\cite{Alzahrani}, and bidder tendency~\cite{Trevathan2008446}. These studies have led to the development of various detection and prevention mechanisms, which can be classified into offline and real-time approaches~\cite{MAJADI20171}.

Initial detection methods relied on reputation-based scoring systems. Trevathan et al. \cite{Trevathan2008446} introduced the Shill Score Algorithm, which tracks six bidding patterns to identify SB. Similarly, Rubin et al. \cite{Rubin2005270} proposed a statistical reputation system that ranks bidders based on anomaly scores. While these solutions are effective for post-auction fraud analysis, they do not provide real-time mechanisms to prevent SB during the auction.

Machine learning techniques have been explored for real-time fraud detection. Anowar et al. \cite{Anowar202081} and Alzahrani et al. \cite{Alzahrani2020269} developed classification models that cluster bidders based on their behaviors in large-scale datasets. Despite their high detection accuracy, these models face scalability issues when applied to real-time bidding environments, where rapid decision-making is required. Hybrid detection approach was proposed by Adabi et al. \cite{Adabi20222381}. it combines offline and real-time mechanisms using genetic algorithms to create favorable conditions for honest bidders. However, this method does not scale well to detect real-time collusion in large-volume auctions.

While most studies focus on detecting SB, fewer works propose mechanisms to prevent it. Kaur et al \cite{Kaur2015381} introduced a variable bid fee model that charges a fixed  2\% fee per bid. 
while this method discourages aggressive bidding, it does not differentiate between honest and fraudulent bidders, leading to unfair penalties for legitimate participants. Komo et al. \cite{komo2024shill} proposed an optimal reserve price strategy to make SB unprofitable. However, this method  limited to resource allocation auctions and does not address bidder collusion in open-bid English auctions.

\subsection*{Blockchain and Auctions} 
Blockchain has been proposed as a secure and transparent solution for online auctions~\cite{Dai2024}. While several blockchain-based auction applications have been explored, limited research has focused on monitoring bidding patterns, particularly in open-outcry auctions.

Most of the studies in the literature have focused on sealed-bid auction applications, Wu et al. \cite{Wu2019} introduced CReam, the first application of smart contracts to deter bid shielding in sealed-bid auctions. Their approach ensures bid confidentiality but does not handle open-bid auction frauds such as SB.  Similarly, Xiong et al. \cite{Xiong2021} designed an anti-collusion smart contract for data auctions, however, their system relies on centralized seller control, reducing decentralization benefits. 

Other studies have focused on enhancing privacy-preserving in blockchain-based auctions. Dai et al. \cite{Dai2024} proposed a sealed-bid auction using zero-knowledge proofs (ZKPs) and multi-party computation (MPC) to prevent bid manipulation while preserving bidder privacy. Guo et al. \cite{Guo2022}, as well introduced a combinatorial auction model for decentralized edge computing using blockchain and differential privacy to ensure bid confidentiality and optimal resource allocation. 

Additional research focused on optimizing blockchain-based auction performance. Omar et al. \cite{omar2021implementing} presented a reverse auction system that extends blockchain use to offer submission through IPFS~\cite{IPFSDocs2025}, reducing reliance on on-chain storage. However, gas costs remain a limitation. To address this, Zhang et al. \cite{Zhang2023} proposed a low-cost verifiable auction mechanism designed to minimize transaction fees and improve cost efficiency.

\subsection{Research Gap and Our Contributions}
Prior research has explored various strategies for SB detection, but the used patterns to identify shill bidders are not always reliable~\cite{MAJADI20171}. Most existing solutions focus on post-auction detection rather than real-time prevention~\cite{Dong2009245}. Blockchain technology offers advantages in transparency, decentralization, and security. However, there is a notable gap in addressing fraud behaviors and mitigating bid manipulation using this technology~\cite{shi2022integration}. Current methods for addressing SB  also face scalability issues related to computational complexity, real-time effectiveness, and cost efficiency~\cite{MAJADI20171}. Their integration into smart contracts introduces additional challenges due to gas fees and storage overhead, which can become significant in high-volume auctions~\cite{shi2022integration}.
Therefore, preventing SB activities within decentralized auctions requires solutions with low computational complexity, support for multiple sellers, and real-time detection. Additionally, they also need to reduce the impact on honest bidders while retaining fair competition in trading, user participation, and realism~\cite{Adabi20222381}.

To address these challenges, we present a blockchain-enabled English auction framework that adopts real-time SB prevention through smart contracts. The main contribution of the paper is a new dynamic penalty mechanism for adjusting bidding fees according to the suggested Bid Shill Score, which examines nine potential suspicious bidding patterns to make shill bidding economically unfeasible. Additionally, we integrate these metrics within the smart contract-based auction mechanism to create a transparent consensus mechanism that discourages collusive bidding while maintaining fair competition.

\section{Blockchain-Based Framework for Shill Bidding Prevention}

The proposed blockchain-based auction framework consists of an architecture that enforces fair bidding practices and a penalty mechanism to deter fraudulent activity. We first present the system architecture, including its components and stakeholder interactions. Next, we introduce the formal mathematical models, including the nine metrics used for SB detection, the Bid Shill Score (BSS), and the penalty fee computation. Finally, we describe the financial impact of The dynamic Penalties on Shill Bidders.

\subsection{System Architecture}

Figure~\ref{fig:arch} illustrates the architecture of the blockchain-based auction system, which employs smart contracts to detect and prevent SB. The system consists of three main layers. The first one is dedicated to the involvement of three main actors: sellers, honest bidders, and shill bidders, all of whom interact with the auction smart contract.

\begin{figure*}[!tb]
\centerline{\includegraphics[width=1\textwidth]{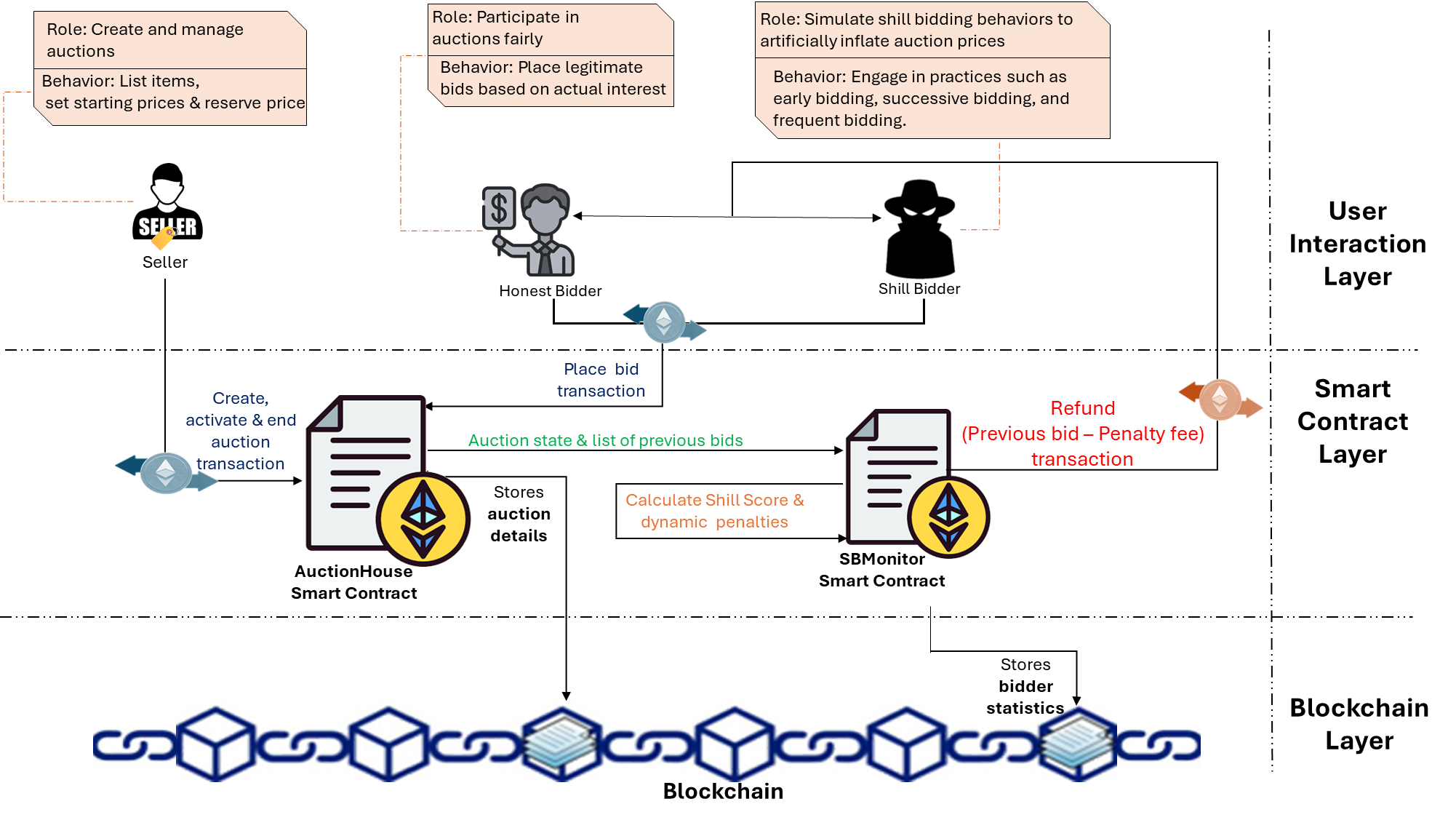}}
\caption{The proposed auction system architecture}
\label{fig:arch}
\end{figure*}

Sellers create and manage auctions by listing items and defining auction parameters. Bidders participate by placing bids, with honest bidders following fair bidding practices and shill bidders attempting to manipulate auction prices.

The smart contract layer consists of two contracts that govern the auction process and enforce bidding rules. The $AuctionHouse$ smart contract handles the entire auction lifecycle, allowing users to create auctions, activate them, place bids, and finalize transactions. The SB patterns monitoring smart contract validates bids by calculating the BSS and applying penalties. This contract continuously monitors bidding behavior to discourage unfair practices while ensuring fair competition. 

Finally, the Blockchain layer integrates the blockchain technology used as the basis for all the supported functionalities.

Figure~\ref{fig:class} illustrates the class diagram of the implemented system, representing the core entities and their relationships within the blockchain-based auction system. The \texttt{Bidder} and \texttt{Seller} classes extend from \texttt{Account}, where bidders participate in auctions and sellers list and manage them. The \texttt{Asset} class represents auctioned items, storing their attributes such as title and description. The \texttt{Auction} class handles auction parameters, including start and end times, current bid, reserve price, and bid increment rules. Sellers list assets in auctions, and bidders place bids through the \texttt{Bid} class, which records bid details such as bidder address, bid amount, and timestamp.

\begin{figure*}[htbp]
\centering
\includegraphics[width=0.91\textwidth]{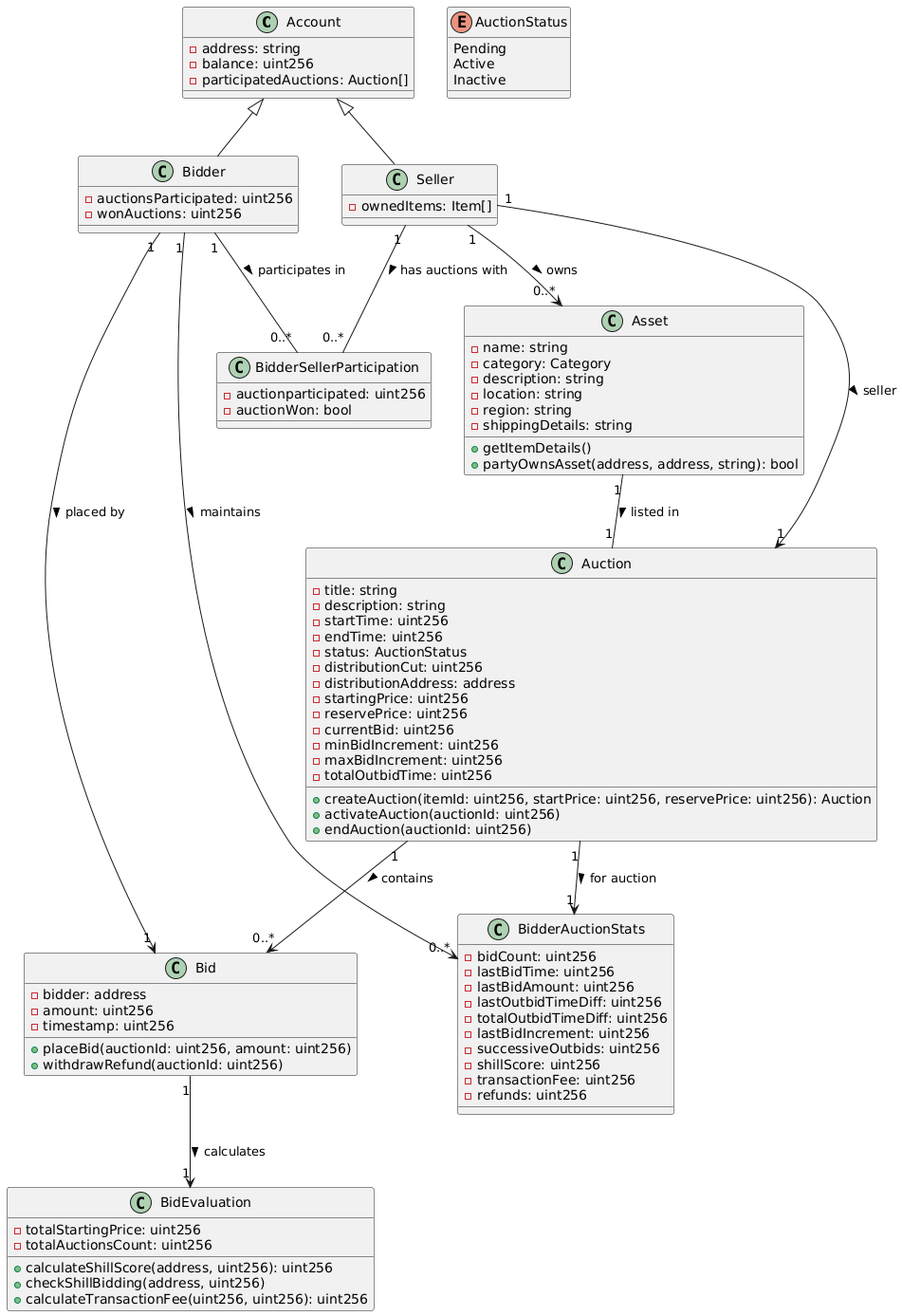}
\caption{Class 
diagram of the blockchain-based auction system}
\label{fig:class}
\end{figure*}

To evaluate bidder behavior, the \texttt{BidderAuctionStats} and \texttt{BidderSellerParticipation} classes track bidding history and patterns, including bid frequency, last bid time, and bidder tendencies toward specific sellers. The \texttt{BidEvaluation} class performs the system's core calculations, including computing the BSS and penalty fees.

Figure \ref{fig:sequence_diagram} presents a sequence diagram illustrating the system's operational flow. The process begins when a seller creates an auction for an asset. The system first verifies the seller's ownership before allowing the auction to be listed. Once confirmed, the seller configures auction parameters such as the starting price, reserve price, and duration. After creation, the seller must activate the auction to enable bidding.
\begin{figure*}[htbp]
\centering
\includegraphics[width=0.8\textwidth]{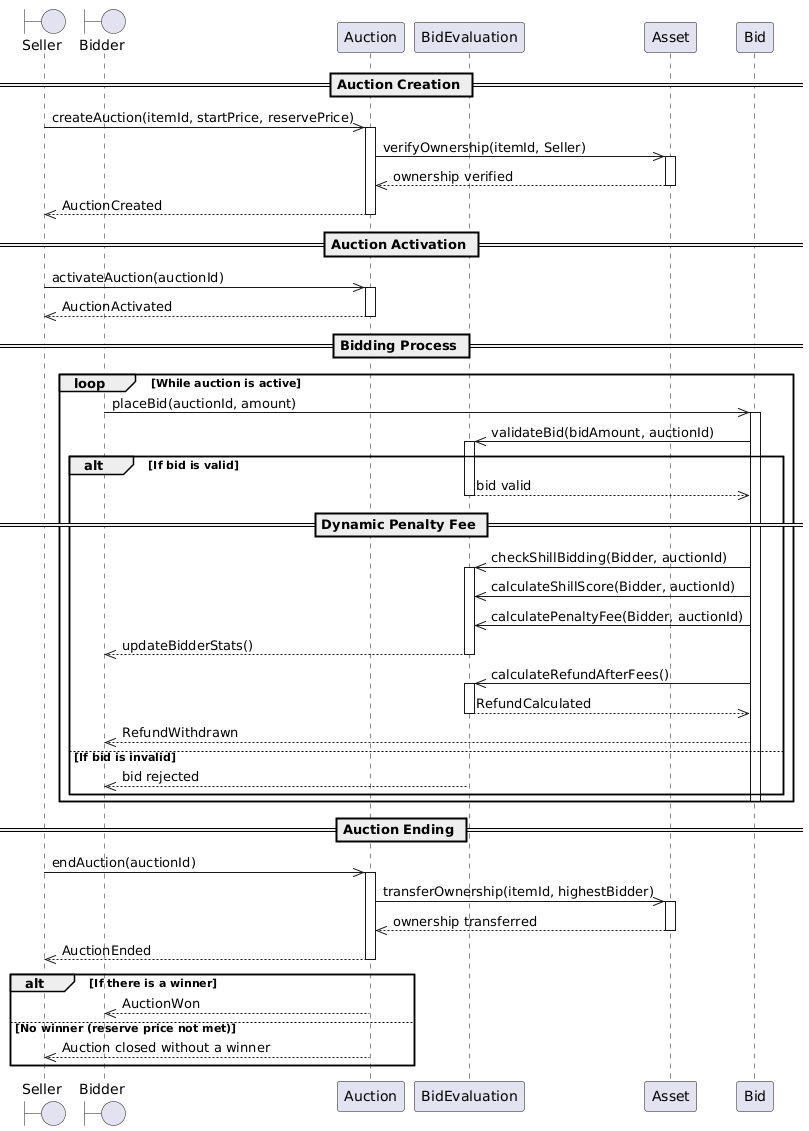}
\caption{Sequence diagram for interactions in the auction process}
\label{fig:sequence_diagram}
\end{figure*}

Once the auction is active, bidders can place bids using the $placeBid$ function. This function validates the bid amount and calls the $processPenalty$ function to compute bid fees for the previous highest bidder. The $processPenalty$ function calculates the dynamic penalty fee using the $calculateShillScore$ function, which determines whether the previous bidder showed SB patterns.

After processing the penalties, the $placeBid$ function refunds the previous highest bidder, deducting the applicable penalty fees from the refunded amount. Finally, when the auction concludes, the seller or contract owner calls the $endAuction$ function to finalize the process. If the highest bid meets or exceeds the reserve price, the smart contract transfers ownership of the asset to the highest bidder, and the seller receives the corresponding payment. However, if the reserve price is not met or no bids are placed, the auction ends without a sale. In this case, the asset is returned to the seller, and any previous highest bidders receive a refund after penalty deductions. 

\subsection{Bid Shill Score Computation and Dynamic Penalty Fee Model}

We propose a dynamic penalty fee model, $F_{i,j,k}$, designed to discourage SB. This bid fee is determined based on the Bid Shill Score (BSS), which identifies nine suspicious bidding patterns previously highlighted by researchers~\cite{MAJADI20171}. The concept of BSS builds on the Shill Score Algorithm (SSA), originally proposed by Trevathan et al.~\cite{Trevathan2008446}, which detects six SB patterns either at the end of an auction or during different auction stages using the Live Shill Score (LSS)~\cite{Majadi2018}.

Building on this work, we extend the detection scope from six to nine SB patterns and apply it dynamically with each bid to compute penalty fees in real time. As a bidder's score increases due to detected suspicious bidding behaviors, the penalty fee proportionally rises. This mechanism discourages SB by making aggressive bidding across multiple accounts financially costly.

Before introducing the metrics, we define the sets and notations used throughout the model.

Let:

\begin{itemize}
    \item \( B \): The set of all bidders participating in the auction system. Each bidder is denoted as \( i \in B \).
    \[
    B = \{i_1, i_2, i_3, \dots, i_n\}
    \]
    Where \( i_1, i_2, \dots, i_n \) represent individual bidders.

    \item \( S \): The set of all sellers that listed auctions in the system. Each seller is denoted as \( s \in S \).
    \[
    S = \{s_1, s_2, s_3, \dots, s_n\}
    \]
    Where \( s_1, s_2, \dots, s_n \) represent individual sellers.

    \item \( A \): The set of all auctions. Each auction is denoted as \( j \in A \). The number of auctions in which bidder \( i \) has participated is donated as \( m_i \), and the number of auctions where bidder \( i \) participated in with a specific seller \( s \) is represented as \( m_i^s \).
    \[
    A = \{j_1, j_2, j_3, \dots, j_m\}
    \]
    Where \( j_1, j_2, \dots, j_m \) represent individual auctions.

    \item \( b_{i,j,k} \): A specific bid placed by bidder \( i \) in auction \( j \) at time \( t_{i,j,k} \), where \( k \) denotes the sequence of bids placed by bidder \( i \) in auction \( j \) and \( t_{j,k-1} \) represents the time of the previous bid in auction \( j \).

    \item \( B_j \): The set of all bids placed in auction \( j \). The size of this set is denoted as \( n_j \). The total number of bids placed by bidder \( i \) in auction \( j \) is donated by \( n_i^j \), and the number of times bidder \( i \) outbids themselves in auction \( j \) is donated \( n_i^{\text{self}, j} \). 
    \[
    B_j = \{b_{j,1}, b_{j,2}, \dots, b_{j,n_i^j}\}
    \] 
    \item \( W_i \): The set of auctions won by bidder \( i \). The size of this set is donated \( w_i \)
    \[
    W_i = \{j_1, j_2, \dots, j_w\}
    \]
    Where \( j_1, j_2, \dots, j_w \) represent the auctions won by bidder \( i \).

    \item \( t_{\text{start}}^j \) and \( t_{\text{end}}^j \): The start and end times of auction \( j \), respectively.
    \item \( \Delta p_i^j \): The difference between consecutive bids placed by bidder \( i \) in auction \( j \).
    \item \( \Delta p_{\text{min}}^j \): The minimum bid increment in auction \( j \).
    \item \( \Delta p_{\text{max}}^j \): The maximum bid increment in auction \( j \).
    \item \( S_j \): The starting price of auction \( j \).
    \item \( P_{\text{avg}} \): The average price of similar items in the market for auction \( j \).
    \item \(\phi = 10000\): A fixed scaling factor used to normalize metric values in Solidity smart contract That allows the representation of percentages with two decimal places of precision.

\end{itemize}

\subsubsection{Patterns Metrics}
To detect SB and calculate the BSS, we use a set of metrics that evaluate bidders' patterns during auctions. These metrics were initially introduced and validated by Trevathan et al.~\cite{Trevathan2008446} and Alzahrani et al.~\cite{Alzahrani}. In this study, we have refined and extended these metrics to better capture fraudulent bidding patterns. Specifically, we introduced the Early Bidding and Late Bidding metrics to capture the tendencies of bidders who place bids more either at the beginning or end of an auction. Furthermore, we propose the Successive Outbidding metric to track how frequently bidders outbid themselves, and the Bidder Tendency metric to identify new accounts or accounts with a high participation rate in auctions from the same seller. These metrics have been adapted for blockchain-based auctions and are designed to be efficiently implemented in smart contracts using Solidity~\cite{solidity2024}.

\begin{itemize}

    \item \textbf{Early Bidding} ($\alpha$): Measures how early a bidder starts bidding in relation to other participants using Equation~\ref{eq:alpha}. Shill bidders often bid very early to influence the auction from the beginning.

        \begin{equation} \label{eq:alpha}
            \alpha_{i,k}^j = 
            \frac{\phi \times (t_{\text{end}}^j - t_{i,j,k})}{t_{\text{end}}^j - t_{\text{start}}^j}
        \end{equation}
    
    \item \textbf{Bid Increment} ($\beta$): Evaluates the increase in bid value made by a bidder in successive bids using Equation~\ref{eq:beta}. Shill bidders tend to place small incremental bids to slowly raise the price without drawing suspicion. 

        \begin{equation} \label{eq:beta}
            \beta_{i,k}^j = \phi - \frac{\phi \times (\Delta p_{i,k}^j - \Delta p_{\text{min}}^j)}{\Delta p_{\text{max}}^j - \Delta p_{\text{min}}^j}
        \end{equation}

    \item \textbf{Outbid Time} ($\gamma$): Captures how quickly a bidder places bids compared to the average outbid time in the auction using Equation~\ref{eq:gamma}. Shill bidders bid much faster than legitimate bidders to create fake competition with honest bidders. 
    
        \begin{equation} \label{eq:gamma}
            \gamma_{i,k}^j = \frac{\phi}{n_i^j} \sum_{k=1}^{n_i^j} \left( \phi - \frac{t_{i,j,k} - t_{j,k-1}}{T_{\text{avg}}^j} \right)
        \end{equation}

        Where:
        \begin{itemize}
         \item \( T_{\text{avg}}^j \): The average outbid time for all bids in auction \( j \), calculated as:
            \[
            T_{\text{avg}}^j = \frac{1}{n_j - 1} \sum_{k=2}^{n_j} (t_{j,k} - t_{j,k-1})
            \]
        \end{itemize}
        
     \item \textbf{Bid Frequency} ($\delta$): Captures how often a bidder places bids in a single auction using Equation~\ref{eq:delta}. A higher bid frequency suggests the bidder is driving up the price without intending to win.
    
        \begin{equation} \label{eq:delta}
            \delta_{i,k}^j = \frac{\phi \times n_i^j}{n_j}
        \end{equation}

    \item \textbf{Bidder Tendency} ($\epsilon$): Measures how frequently a bidder participates in auctions held by the same seller using Equation~\ref{eq:epsilom}. Bidders with a high frequency of participation in the same seller's auctions are more likely to be shill bidders.
    
        \begin{equation} \label{eq:epsilom}
            \epsilon_i = \frac{\phi \times (m_i^s - w_i^s)}{m_i}
        \end{equation}
    
    \item \textbf{Winning Ratio} ($\zeta$): Tracks how often a bidder loses auctions in which he participates, using Equation~\ref{eq:zeta}. Shill bidders avoid winning auctions to not incur the cost of winning.
        
        \begin{equation} \label{eq:zeta}
        \zeta_i = \phi - \frac{\phi \times w_i}{m_i}
        \end{equation}

    \item \textbf{Successive Outbidding} ($\eta$): Measures how often a bidder outbids their own bid, which is a common indicator of SB. This is calculated using Equation~\ref{eq:eta}.
       
        \begin{equation} \label{eq:eta}
            \eta_{i,k}^j = \frac{\phi \times n_i^{\text{self}, j}}{n_i^j}
        \end{equation}

    \item \textbf{Auction Starting Price} ($\kappa$): indicates how the initial auction price relates to previous similar auctions as shown in Equation~\ref{eq:kappa}. Auctions with low starting prices are more susceptible to SB as they attract more bidders and allow greater price manipulation.
        \begin{equation} \label{eq:kappa}
            \kappa_j = \phi - \frac{\phi \times S_j}{P_{\text{avg}}}
        \end{equation}

      \item \textbf{Late Bidding} ($\iota$): Captures how late a bidder places his bid relative to other participants using Equation~\ref{eq:iota}. Shill bidders stop placing bids before the last stage of the auction duration. In the other hand,
      Strategic last-moment bidding, or sniping, is common sign for legitimate bidders who aim to secure the auction closer to the end.
        
        \begin{equation} \label{eq:iota}
        \iota_{i,k}^j = \frac{\phi \times (t_{i,j,k} - t_{\text{start})}}{t_{\text{end}}^j - t_{\text{start}}^j}
        \end{equation}
    
\end{itemize}

These metrics are designed to capture a broad range of suspicious behaviours, making it difficult for attackers to evade detection without substantially altering their strategies. For instance, Bidder Tendency and Winning Ratio identify repeated interactions with specific sellers, including cases involving multiple accounts. In parallel, Bid Frequency, Bid Increment, Outbid Time, and Successive Outbidding highlight coordinated bidding patterns and timing-based collusion.

\subsubsection{The Bid Shill Score}

After calculating the nine metrics $(\alpha, \beta, \gamma, \delta, \epsilon, \zeta, \eta, \kappa, \iota)$,the $BSS_{i,j,k}$ for a bid $b_{i,j,k}$ is calculated using Equation~\ref{eq:bss}. The weights assigned to each metric in this equation are $w_1 = 8, w_2 = 5, w_3 = 4,w_4 = 2, w_5 = 7, w_6 = 8, w_7 = 7, w_8 = 3, w_9 = 2$, where $n = 9$ represents the total number of suspicious behaviors considered in the model. These weights were determined through experimentation and simulated auctions, as proposed by Trevathan et al. \cite{Trevathan2008446} and Sadaoui et al. \cite{Alzahrani}, and further validated through additional simulations by us. 

The assignment of these weights was guided by the direct impact of each pattern on shill bidding behavior. Metrics strongly related to seller collusion or manipulation (such as Bidder Tendency, Successive Outbidding, and Winning Ratio) were assigned higher weights. In contrast, patterns that may also appear in honest aggressive bidding (such as Late Bidding and Auction Starting Price) received lower weights to avoid false positives.
This deterministic weight configuration aligns with smart contract requirements, ensuring verifiability and consistent application of penalties on-chain.  the $BSS_{i,j,k}$ value ranges between 0 and 100, where a higher BSS indicates a higher likelihood that the bidder is engaging in SB.

\begin{align} \label{eq:bss}
BSS_{i,j,k} &= 2000 \times \frac{\left( w_1 \alpha_i + w_2 \beta_i^j + w_3 \gamma_i + w_4 \delta_i^j + w_5 \epsilon_i^j + w_6 \zeta_i^j + w_7 \eta_i^j + w_8 \kappa_j - w_9 \iota_i^j \right)}{n}
\end{align}

The current model uses fixed weights to ensure transparency and consistency, which aligns with the deterministic nature of smart contracts. This design allows the penalty mechanism to function as a consensus layer, with rules agreed upon at the beginning of the auction. As a result, participants can independently verify that penalties are correctly computed and enforced, without relying on any off-chain logic or subjective interpretation.

\subsubsection{The Dynamic Penalty Fees Calculation}

Once the shill score $BSS_{i,j,k}$ is determined, the percentage of the penalty fee is 
The \textit{dynamic penalty fee} $F_{i,j,k}$for a bid $b_{i,j,k}$ is calculated as a percentage of the bid amount. This percentage increases as the shill score rises, following Equation~\ref{eq:bid_penalty}, where $P_{\text{base}}$ set at 2\%, which is the base percentage of the bid amount that is applied to all participants. This approach is inspired by the work of Kaur et al. \cite{Kaur2015381}. Additionally, The maximum penalty percentage $P_{\text{max}}$ can base up to 5\% of the bid amount, depending on the bidder's shill score.

\begin{equation} \label{eq:bid_penalty}
    F_{i,j,k} = \frac{b_{i,j,k}}{\phi} \times \left( P_{\text{base}} + \frac{BSS_{i,j,k} \times P_{\text{max}}}{\phi} \right)
\end{equation}

\subsubsection{Financial Impact of Dynamic Penalties on Shill Bidding}

SB is primarily motivated by the potential for higher profits than the reserve price \( P_{\text{reserve}} \), which represents the minimum acceptable price set by the seller. By placing fake bids, a seller manipulates the auction to drive up the final price \( P_{\text{final}} \). The potential profit a seller can achieve from SB, before accounting for any costs, is given by Equation \ref{eq:profit}:

\begin{equation} \label{eq:profit} P_{\text{profit}} = P_{\text{final}} - P_{\text{reserve}} \end{equation}

However, in our proposed system, shill bidders $S$ incur costs due to transaction fees $T_{\text{fee}, i,j,k}$ and penalties $F_{i,j,k}$. The total cost \( C_{\text{total}} \) accounts for both penalties applied to fraudulent bids and the gas fees required to execute these transactions. This cost is defined in Equation \ref{eq:total_cost}.

\begin{equation} \label{eq:total_cost} C_{\text{total}} = \sum_{i \in S} \sum_{k=1}^{n_i^j} \left( F_{i,j,k} + T_{\text{fee}, i,j,k} \right) \end{equation}

If \( C_{\text{total}} \) exceeds the potential profit \( P_{\text{profit}} \), then engaging in SB becomes economically unfeasible. The net profit \( R_{\text{profit}} \),
defined in Equation \ref{eq:net_profit}, determines whether the seller benefits from fraudulent activities or incurs a financial loss.

\begin{equation} \label{eq:net_profit} R_{\text{profit}} = P_{\text{profit}} - C_{\text{total}} \end{equation}

\lstdefinelanguage{Solidity}{
  keywords={uint256, function, internal, view, returns, if, else, return, emit, require, public, external},
  keywordstyle=\color{blue}\bfseries,
  ndkeywords={contract, import, pragma, solidity},
  ndkeywordstyle=\color{darkgray}\bfseries,
  identifierstyle=\color{black},
  sensitive=false,
  comment=[l]{//},
  commentstyle=\color{purple}\ttfamily,
  stringstyle=\color{red}\ttfamily,
  morestring=[b]',
  morestring=[b]"
}

\lstset{
   language=Solidity,
   backgroundcolor=\color{lightgray},
   extendedchars=true,
   basicstyle=\footnotesize\ttfamily,
   showstringspaces=false,
   showspaces=false,
   numbers=left,
   numberstyle=\footnotesize,
   numbersep=2pt,
   tabsize=2,
   breaklines=true,
   showtabs=false,
   captionpos=b
}

\section{Implementation details}
Here, we present the implementation of the blockchain-based online auction system with SB prevention. 
The $AuctionHouse$ smart contract serves as the auctioneer, managing the entire auction lifecycle while ensuring transparency and fairness without reliance on a trusted third party. The $SBMonitor$ smart contract tracks bidding behaviour, computes the $BSS$, and enforces penalties on bidders. These smart contracts are deployed on the Ethereum blockchain to guarantee secure and immutable enforcement of auction rules.

The main functionalities of these smart contracts, illustrated in Figure \ref{fig:sequence_diagram}, follow a structured workflow:

\begin{enumerate}
    \item The seller creates and activates an auction.
    \item Bidders submit bids within the auction duration.
    \item The smart contract processes penalties by computing the $BSS$, storing the penalty value, and refunding the previous highest bidder after deducting the penalty fee.
\end{enumerate}

The implementation focuses on the core functions related to SB prevention, including: \textbf{placeBid}, \textbf{processPenalty}, \textbf{calculateShillScore}, and \textbf{calculateTransactionFee}.

The $placeBid$ function, shown in Listing \ref{lst:placeBid}, 
validates first the bid amount and ensures the bidder is not the auction owner. It also confirms that the auction remains active.
    
     \begin{lstlisting}[caption={Solidity Function for Placing a Bid}, label={lst:placeBid}]
    function placeBid(uint auctionId) external payable onlyLive(auctionId) onlyNotOwner(auctionId) auctionExists(auctionId) returns (bool) {
        uint256 amount = msg.value;
        Auction storage a = auctions[auctionId];
        BidderStats storage stats = bidderStats[msg.sender];
    
        if (a.currentBid >= amount) {
            revert BidNotHighEnough("Current bid is not high enough");
        }
    
        if (stats.auctionStats[auctionId].bidCount == 0) {
            stats.auctionsParticipated++;
        }
    
        // Update auction's current bid and add new bid
        a.currentBid = amount;
        a.bids.push(Bid({
            bidder: msg.sender,
            amount: amount,
            timestamp: block.timestamp
        }));
    
        // Call the processPenalty function for the previous bidder
        processPenalty(msg.sender, auctionId, amount, bidIdx, a);
        
        emit BidPlaced(auctionId, msg.sender, amount);
        return true;
    }
    \end{lstlisting}

As bids are placed, the $processPenalty$ function, shown in Listing \ref{lst:processPenalty}, executes three steps. First, it calls $calculateShillScore$ to compute the bidder's $BSS_{i,j,k}$. Next, \textit{calculateTransactionFee} determines the dynamic penalty fee $F_{i,j,k}$ based on the $BSS$ value. Finally, the penalty fee is deducted from the refundable deposit, and the remaining balance is returned to the previous highest bidder.

    \begin{lstlisting}[caption={Solidity Function for Processing Penalty}, label={lst:processPenalty}]
    function processPenalty(address bidder, uint256 auctionId, uint256 amount, uint256 bidIdx, Auction storage a) internal {
        BidderStats storage stats = bidderStats[bidder];
        uint256 shillScore = calculateShillScore(bidder, auctionId);
        uint256 transactionFee = calculateTransactionFee(amount, shillScore);
    
        // Update bidder statistics
        stats.auctionStats[auctionId].shillScore = shillScore;
        stats.auctionStats[auctionId].transactionFee = transactionFee;
    
        // Refund the previous highest bidder
        if (bidIdx > 0) {
            Bid storage previousBid = a.bids[bidIdx - 1];
            uint256 refundAmount = previousBid.amount - (previousBid.amount * previousFee);
            refunds[previousBid.bidder] += refundAmount;
            emit RefundProcessed(previousBid.bidder, refundAmount, stats.auctionStats[auctionId].transactionFee);
        }
    }
    \end{lstlisting}

The data used to compute the $BSS$ and $F_{i,j,k}$ is structured and stored on the blockchain, ensuring transparency and efficient retrieval of bidder activity. The $BidderStats$ and \textit{BidderAuctionStats} structs, shown in Listing \ref{lst:bidderstruct}, define the relationship between a bidder and their bidding history. They track participation and behaviour across auctions, including bid frequencies, outbidding patterns, and interactions with sellers.
 
    \begin{lstlisting}[caption={Solidity Structs for Bidder and Status},label={lst:bidderstruct}]
    
    struct BidderStats {
        uint256 auctionsParticipated;
        uint256 wonAuctions;
        mapping(uint256 => BidderAuctionStats) auctionStats; // auctionId => BidderAuctionStats
    }
    
    struct BidderAuctionStats {
        uint256 bidCount;
        uint256 lastBidTime;
        uint256 lastBidAmount;
        uint256 lastOutbidTimeDiff;
        uint256 totalOutbidTimeDiff;
        uint256 lastBidIncrement;
        uint256 successiveOutbids;
        uint256 shillScore;
        uint256 transactionFee;
    }
    \end{lstlisting}

The mappings in Listing \ref{lst:mappings} represent the relationships between auctions, sellers, and bidders. This structure enables efficient retrieval of bidder data without loops, reducing gas costs during contract execution. In Ethereum-based smart contracts, gas costs determine execution and deployment expenses, as each computational step consumes gas units. Since one of our main objectives is to optimize the cost of running the auction system, minimizing gas consumption is essential. Using mappings instead of arrays reduces unnecessary computations, lowering transaction fees and improving efficiency and scalability \cite{alchemy_solidity_mapping}.

\begin{lstlisting}[caption={Solidity Mappings for Auction and Bidder Tracking}, label={lst:mappings}]
mapping(address => uint256[]) public auctionsRunByUser;
mapping(address => uint256[]) public auctionsBidOnByUser;

mapping(address => mapping(address => uint256)) public sellerBidderParticipations;

// Mapping to track whether a bidder has participated in a specific auction with a seller
mapping(address => mapping(address => mapping(uint256 => bool))) public bidderParticipatedInAuctionWithSeller; // bidder => seller => auctionId => bool

// Mapping to track the number of auctions won by a bidder with a specific seller
mapping(address => mapping(address => uint256)) public bidderAuctionsWonWithSeller; // bidder => seller => count
\end{lstlisting}

\subsection{Testing}
The system was tested and validated using Ganache Truffle Suite (v2.7.1)~\cite{trufflesuiteGanache} to ensure that its logic functions correctly, maintaining fairness and security in the auction process. The system was evaluated by executing the main smart contract functions under various scenarios.

\begin{enumerate}
    \item \textbf{Auction Creation and Activation}\\
    This test verifies the creation and activation of an auction. The seller initiates the auction by setting parameters such as the starting price, reserve price, and duration before enabling bidding. As shown in Figure \ref{fig:create-auct}, an auction with ID 0 was successfully created and activated, with a transaction cost of 0.0083 ETH for creation and 0.0012 ETH for activation.

    \begin{figure*}[h]
        \centering
        \includegraphics[width=0.75\linewidth]{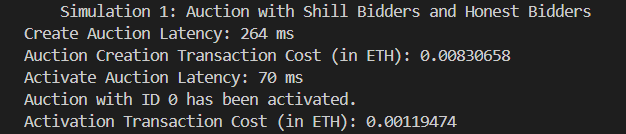}
        \caption{Create and activate auction output details}
        \label{fig:create-auct}
    \end{figure*}
    
    \item \textbf{Bid Submission}\\
    Once the auction is active, bidders can place bids. This test evaluates the $placeBid$ function, ensuring correct bid processing and real-time computation of $BSS$ and $F_{i,j,k}$. As shown in Figure \ref{fig:placeBid}, Bidder $0x5e92F4e4F0b1F2a03F38cDcFCC8f68D808Ac6f5$ placed the ninth bid. The contract tracks the bidder's patterns and applies the calculated penalty.
    
    \begin{figure*}[h]
        \centering
        \includegraphics[width=0.75\linewidth]{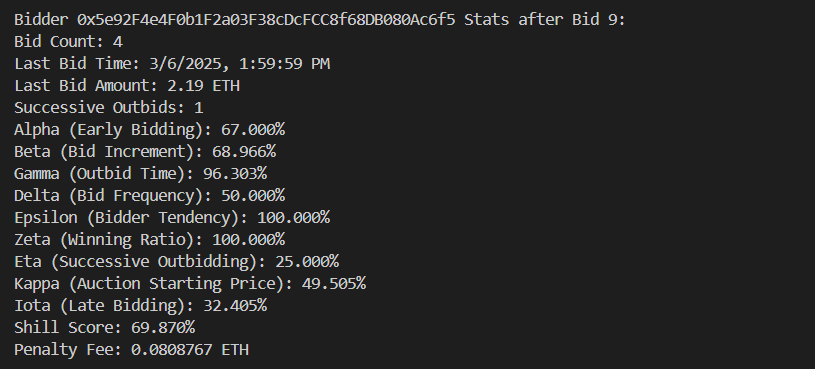}
        \caption{$placeBid$ function output details}
        \label{fig:placeBid}
    \end{figure*}

     \item \textbf{Auction Termination}\\
    The $endAuction$ function ensures that the auction closes once the predefined duration has elapsed, preventing further bids. As shown in Figure \ref{fig:end-auction}, the function successfully finalises the auction based on the highest valid bid.
        
    \begin{figure*}[h]
        \centering
        \includegraphics[width=0.75\linewidth]{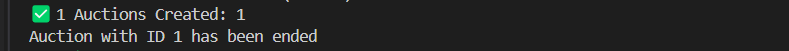}
        \caption{$endAuction$ function output details}
        \label{fig:end-auction}
    \end{figure*}
    
\end{enumerate}

\section{Discussion}
The decision to implement SB prevention through smart contracts ensures a transparent, decentralized, and tamper-proof mechanism. Smart contracts serve as a consensus layer between bidders and sellers, enforcing predefined rules without requiring a trusted third party. This makes fraudulent auction manipulation unprofitable for dishonest sellers while discouraging malicious behavior through automated penalties. Additionally, smart contracts manage assets, handle deposits, and securely store transaction history in a decentralized environment, making them well-suited for addressing auction fraud \cite{Xiong2021}.

Shill bidders intend to inflate auction prices beyond the reserve to boost seller profits.
To evaluate the robustness of the proposed BSS model, simulations were conducted to replicate a variety of advanced manipulation strategies observed in online auctions. These included single-account SB, multi-account coordination, and time-based collusion. Each scenario was executed over 50 independent auction runs to evaluate consistency and detection accuracy.
\\
Table~\ref{tab:attack_bss_metrics} summarises the manipulation tactics and indicates which BSS metrics were most affected in each case. The results confirm that the model captures distinct behavioural patterns associated with different forms of SB.
\\
\begin{table}[h]
    \centering
    \caption{Mapping of BSS Metrics Impact for Different SB Strategies}
    \label{tab:attack_bss_metrics}
    \begin{tabular}{p{3.5cm} p{2.5cm} p{2.5cm} p{2.5cm}}
        \hline
        \textbf{BSS Metric} & \textbf{Single-Account SB} & \textbf{Multi-Account SB} & \textbf{Time-Collusion SB} \\
        \hline
        $\alpha$          & High   & Medium & Low    \\
        $\beta$           & Medium & Medium & High   \\
        $\delta$          & High   & High   & High   \\
        $\epsilon$      & High   & High   & Medium \\
        $\gamma$            & Medium & Medium & High   \\
        $\eta$    & Medium & High   & High   \\
        $\kappa$ & Low    & Low    & Low    \\
        $\zeta$           & High   & High   & Medium \\
        $\iota$            & Low    & Low    & High   \\
        \hline
    \end{tabular}
\end{table}
\\
The results show a clear distinction between the manipulation strategies. As illustrated in Figure~\ref{fig:security_analysis_sim}, the median BSS reached approximately 76\% for multi-account SB, 72\% for single-account shilling, and 68\% for time-based collusion. In contrast, honest participants—including aggressive but fair bidders—consistently displayed substantially lower BSS values. These stable outcomes across multiple runs indicate the system’s capacity to distinguish legitimate competitive behaviour from manipulative bidding. The scoring also remained consistent and reproducible, with repeated simulations showing minimal variance, confirming the reliability of the detection mechanism under varied auction conditions.
\\
\begin{figure}[htbp]
\centering
\includegraphics[width=0.6\linewidth]{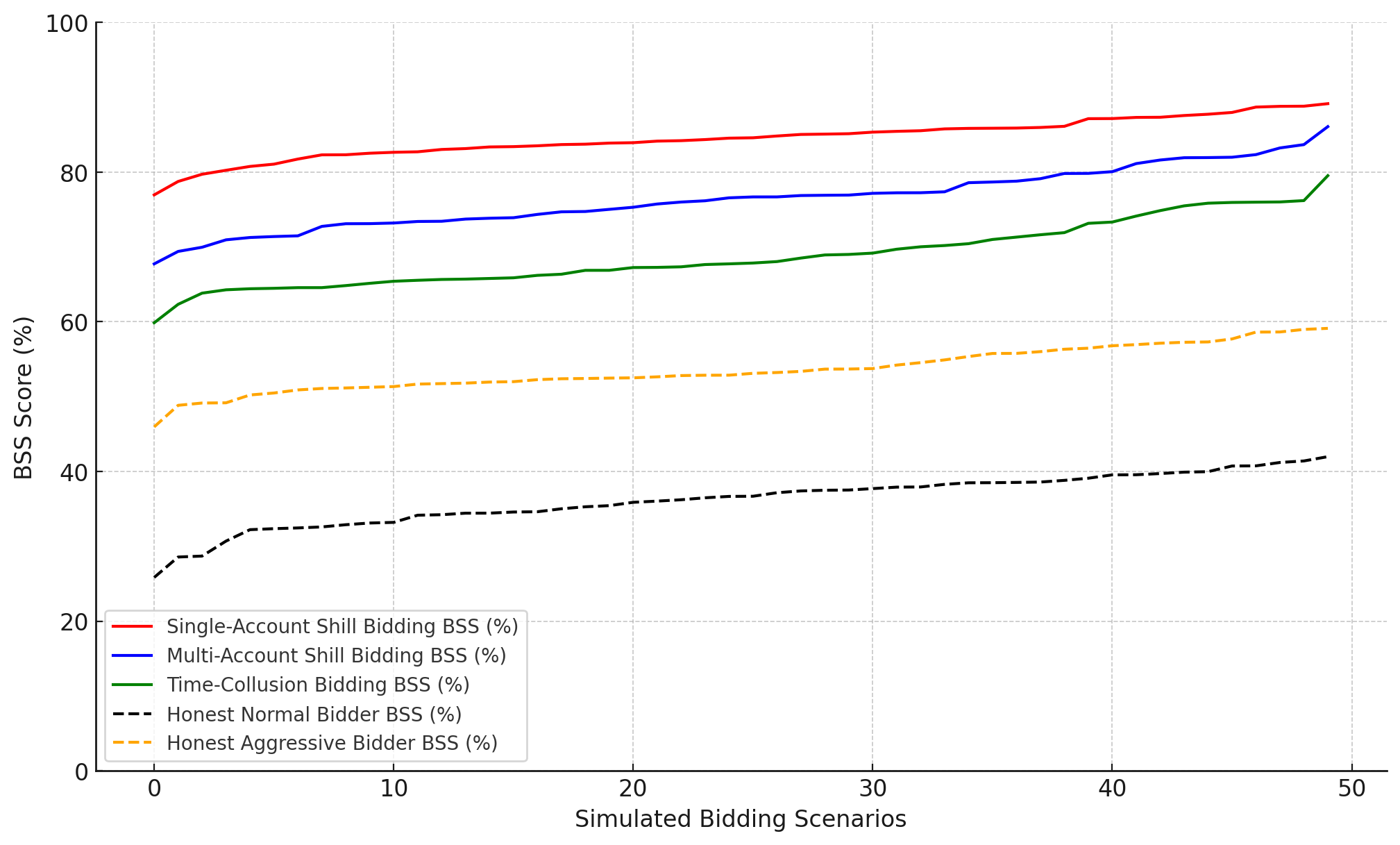}
\caption{BSS results across 50 auction simulations for Single-Account, Multi-Account, and Time-Collusion SB strategies.}
\label{fig:security_analysis_sim}
\end{figure}
\\
Beyond simulated testing, the system was also evaluated in a controlled auction scenario reflecting real bidding patterns involving both honest and shill bidders. Table~\ref{tab:bid_info} presents a case where two shill bidders were actively engaged to inflate the auction price above the reserve. Using separate accounts, the malicious participants executed successive bids characterized by early bidding, rapid outbidding intervals, and incremental price steps. The reserve price was set at 2 ETH, and the auction progressed with layered SB actions until honest bidders intervened.

\begin{table}[h]
    \centering
    \caption{Bid history of auction test with shill bidders}
    \label{tab:bid_info}
    \begin{tabular}{l c r r r}
        \hline
        Bidder Address &Bid Time & Bid amount (ETH) & Shill Score (\%) & Penalty \\
        \hline
        ShillBidder1 &  10:43:32 & 0.40 & 83.76 & 0.0176 \\
        ShillBidder1 &  10:43:59 & 0.50 & 84.10 & 0.022 \\
        ShillBidder2 &  11:30:06 & 0.70 & 50.83 & 0.012 \\
        ShillBidder2 &  11:30:35 & 1.00 & 60.42 & 0.032 \\
        ShillBidder2 &  11:30:49 & 1.30 & 67.70 & 0.047 \\
        ShillBidder2 &  11:31:02 & 1.40 & 81.83 & 0.06 \\
        HonestBidder1 &  11:38:32 & 1.45 & 43.00 & 0.034 \\
        ShillBidder2 &  11:41:54 & 1.55 & 76.98 & 0.063 \\
        HonestBidder2 &  11:42:42 & 1.71 & 45.66 & 0.042 \\
        HonestBidder1 &  11:59:28 & 1.72 & 42.82 & 0.04 \\
        HonestBidder2 &  11:59:30 & 1.77 & 57.77 & 0.055 \\
        ShillBidder1 &  13:26:17 & 1.87 & 69.26 & 0.068 \\
        HonestBidder1 &  13:29:38 & 1.91 & 37.12 & 0.039 \\
        ShillBidder1 &  13:30:00 & 2.21 & 57.21 & 0.068 \\
        ShillBidder1 &  13:30:16 & 2.31 & 67.31 & 0.082 \\
        ShillBidder1 &  13:30:44 & 2.51 & 65.15 & 0.087 \\
        ShillBidder2 &  14:11:10 & 2.81 & 54.25 & 0.082 \\
        HonestBidder2 &  14:13:55 & 2.94 & 45.31 & 0.072 \\
        ShillBidder1 &  14:14:13 & 3.04 & 64.86 & 0.11 \\
        HonestBidder1 &  14:14:25 & 3.06 & 35.47 & 0.06
\\\hline
    \end{tabular}
\end{table}

We compare the seller’s profit before and after applying penalties. Without penalties, the seller's net profit $P_{\text{profit}}$, as defined in Equation~\ref{eq:net_profit}, would be 1.06 ETH. After applying $F_{i,j,k}$, the total cost $C_{\text{total}}$ charged to shill bidders, calculated using Equation~\ref{eq:total_cost}, is 0.85 ETH. Consequently, the seller’s actual net profit $R_{\text{profit}}$ is reduced to 0.21 ETH, which is only 10\% higher than the reserve price.

The observed reduction in profit removes the financial motivation for SB, making continued manipulation economically unviable. In the simulated auction, two coordinated shill bidders employed a range of deceptive tactics, including early bidding, successive outbidding, short bid intervals, and bid layering across two separate accounts. These behaviours were reflected in the recorded bidding sequence and were successfully identified by the system. The ability to detect both timing-based and identity-obscured manipulations confirms that the combined metric model can reliably capture advanced collusion strategies. These results indicate that the system can recognise not only isolated suspicious signals but also coordinated patterns consistent with real-world auction fraud.

Compared to earlier methods summarised in Table~\ref{tab:comparison}, previous solutions have largely focused on offline or post-auction detection within centralised environments. These approaches are susceptible to data tampering and lack transparency. In contrast, the proposed system enforces detection directly on-chain, enabling real-time monitoring in English auctions. The open bidding process is well aligned with blockchain’s transparency, allowing bids to be evaluated as they are submitted.
Prior work, such as Trevathan’s Shill Score Algorithm~\cite{Majadi2018} and Rubin’s anomaly-based reputation system~\cite{Rubin2005270}, operate only after the auction concludes. By contrast, our system introduces deterrence during the bidding process itself. Machine learning approaches~\cite{Anowar202081} offer strong detection capabilities, but face scalability challenges in real-time contexts and rely on stochastic processes, which are incompatible with the deterministic execution required by smart contracts. Kaur’s fixed-fee scheme~\cite{Kaur2015381}, while simple, lacks adaptability—its static penalties may unfairly penalise legitimate bidders.
The proposed model addresses these limitations by introducing dynamic penalties that scale with the degree of suspicious behaviour, ensuring fair treatment for competitive but honest participants. It is implemented using lightweight logic and efficient storage, tailored for smart contract environments. By combining tamper resistance, real-time enforcement, and economic disincentives, the system provides a reliable approach for fraud detection in open decentralised auctions. Additionally, all interactions are recorded on-chain, supporting retrospective analysis and enabling the smart contract to function as a consensus mechanism between participants.

\begin{table}[htbp]
\centering
\setlength{\tabcolsep}{6pt}
\caption{Comparison of SB Techniques}
\renewcommand{\arraystretch}{1.4}
\resizebox{\textwidth}{!}{%
\begin{tabular}{p{3.2cm} p{2.2cm} p{2.5cm} p{3.2cm} c p{3cm} c p{2.8cm} }
\hline
\textbf{Paper} & \textbf{System Type} & \textbf{Prevention or Detection} & \textbf{Technique} & \textbf{Real-time} & \textbf{Data Source} & \textbf{Scalability Challenge} & \textbf{Cost Sensitivity} \\
\hline

Majadi et al.~\cite{Majadi2018} & Centralized & Detection & Live Shill Score Algorithm & Partial & Synthetic + Real & \xmark & \xmark \\
Anowar et al. \cite{Anowar202081} & Centralized & Detection & Machine learning & \xmark & Synthetic & \checkmark & \xmark \\
Kaur et al.~\cite{Kaur2015381} & Centralized & Prevention & Variable Bid Fee & \checkmark & Not Required & \xmark & \xmark \\
Our Work & Decentralized & Prevention & Smart Contract + Dynamic Penalty & \checkmark & Synthetic + Real & \checkmark & \checkmark (Gas fees) \\

\hline
\end{tabular}%
}
\label{tab:comparison}
\end{table}

As this represents the first approach to preventing SB in decentralized auctions using smart contracts, further validation is required. In particular, testing must be conducted with actual blockchain-based auction data to evaluate the effectiveness of the nine selected metrics and to standardize their application in identifying suspicious bidders.

While Solidity mappings and event logs improve data access efficiency, gas fees and storage overhead remain concerns, particularly in high-volume scenarios.
To assess the gas cost implications of the proposed system, we conducted an experiment measuring both gas consumption and execution latency for core smart contract operations, including auction creation, activation, bidding, and finalisation. Multiple iterations were performed, with specific attention to the performance impact introduced by the SB detection and dynamic penalty logic embedded within the \texttt{placeBid} function.
Figure~\ref{fig:cost_metrics} presents the results of this evaluation, conducted in a simulated environment using Ganache. As shown in Figure~\ref{fig:fungas}, gas consumption for \texttt{placeBid} increases linearly with the number of bids, reaching approximately 35,000 gas units for auctions involving 100 bids. This reflects the cost of storing bidder histories and performing real-time behavioural checks on-chain. Although mappings and event logs reduce redundancy and improve data retrieval, maintaining detailed records entirely on-chain remains expensive. Nevertheless, gas usage for \texttt{placeBid} remains within a practical range, especially when compared to more resource-intensive functions such as \texttt{createAuction}. This suggests that while the system adds complexity to support secure and fair bidding, the overhead remains acceptable for typical workloads.
This highlights a broader limitation of the current architecture, which stores all bid data, bidder statistics, and auction records on-chain to ensure transparency and verifiability. To quantify the storage impact, we conducted an estimation based on the smart contract’s data structures, as summarised in Table~\ref{tab:storage_reduction}. For a scenario involving 1,000 auctions with 100 bids each, the estimated on-chain storage requirement is approximately 13.3~MB. This would incur considerable gas costs and present scalability challenges.
To mitigate this issue, a hybrid storage approach could be adopted. By retaining only essential data—such as the final bid, bidder address, and verification hashes—on-chain, and offloading detailed bidding histories and statistics to decentralised storage platforms such as IPFS, the on-chain footprint can be reduced by over 90\%. This makes the system more suitable for large-scale, high-frequency auction settings.
\\
\begin{table}[h]
    \centering
    \caption{Estimated On-Chain Storage vs. Hybrid Storage and Reduction}
    \label{tab:storage_reduction}
    \resizebox{\textwidth}{!}{%
    \begin{tabular}{l r r r}
        \hline
        Component & Current On-Chain Size & Hybrid On-Chain Size & Reduction \\
        \hline
        Auction & ~448 bytes & ~300 bytes & ~33\% \\
        Bid (100 bids) & ~8,400 bytes & ~100 bytes (hash) & ~99\% \\
        BidderAuctionStats (10 bidders) & ~4,480 bytes & ~200 bytes & ~95\% \\
        \hline
        Total per auction & ~13,300 bytes (13.3 KB) & ~600 bytes & ~95\% \\
        Total for 1,000 auctions & ~13.3 MB & ~0.6 MB & ~95\% \\
        \hline
    \end{tabular}
    }
\end{table}
\\
To assess potential latency concerns associated with deploying the BSS model on-chain and computing nine metrics per bid, we evaluated the system’s performance under increasing load. Figure~\ref{fig:funcLatency} shows the average transaction latency for key contract functions across varying numbers of bids. The results indicate that the \texttt{placeBid} function maintains stable latency, averaging around 300~ms even with 100 bids, despite the additional overhead introduced by behaviour analysis and metric computation. This remains within acceptable bounds for decentralised auction environments. For context, most eBay auctions receive fewer than 50 bids~\cite{5070742}, with the Palm Pilot PDA dataset reporting a maximum of 51 bids~\cite{ebay_palm_pilot_dataset}.
Figure~\ref{fig:latency} compares the latency of the proposed function with a baseline version. While the dynamic function introduces additional delay due to real-time score computation, it scales linearly and reaches approximately 200~ms at 100 bids, compared to 30~ms for the baseline. A parallel stress test using 100 accounts (Figure~\ref{fig:scalability}) shows peak latency of 120~ms, confirming that the penalty mechanism does not cause excessive delay or compromise responsiveness. Even under high load, the bidding process remains responsive and does not show signs of slowdown or vulnerability to denial-of-service behaviour.
It is important to note, however, that these results were obtained in a controlled environment using the Ganache network, which does not fully replicate the characteristics of live Ethereum testnets or mainnet. Real-world deployments are subject to greater variability in gas prices, network congestion, and block confirmation times. Future testing on public testnets will be conducted to better capture these operational conditions.

\begin{figure}
    \centering
    \begin{subfigure}[b]{0.48\textwidth}
        \centering
        \includegraphics[width=\textwidth]{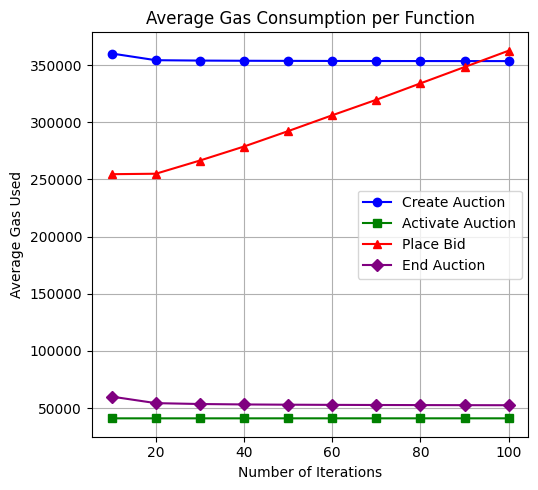}
        \caption{Gas cost of the different system functionalities}
        \label{fig:fungas}
    \end{subfigure}
    \begin{subfigure}[b]{0.48\textwidth}
        \centering
        \includegraphics[width=\textwidth]{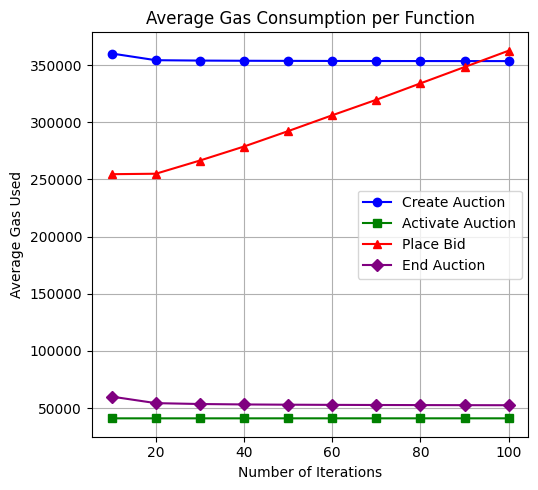}
        \caption{Time cost of functions across different experiments}
    \label{fig:funcLatency}
    \end{subfigure}

    \begin{subfigure}{0.48\textwidth}
        \centering
        \includegraphics[width=\textwidth]{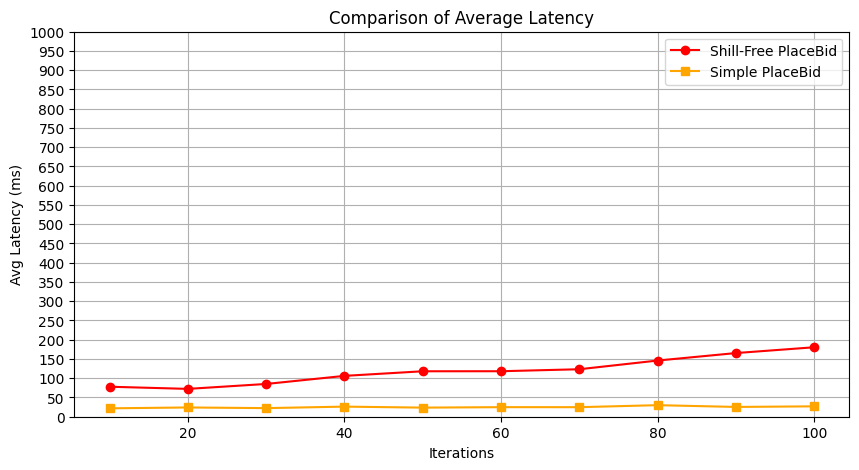}
        \caption{Latency of place-bid function with and without shill score}
        \label{fig:latency}
    \end{subfigure}
    \begin{subfigure}{0.48\textwidth}
        \centering
        \includegraphics[width=\textwidth]{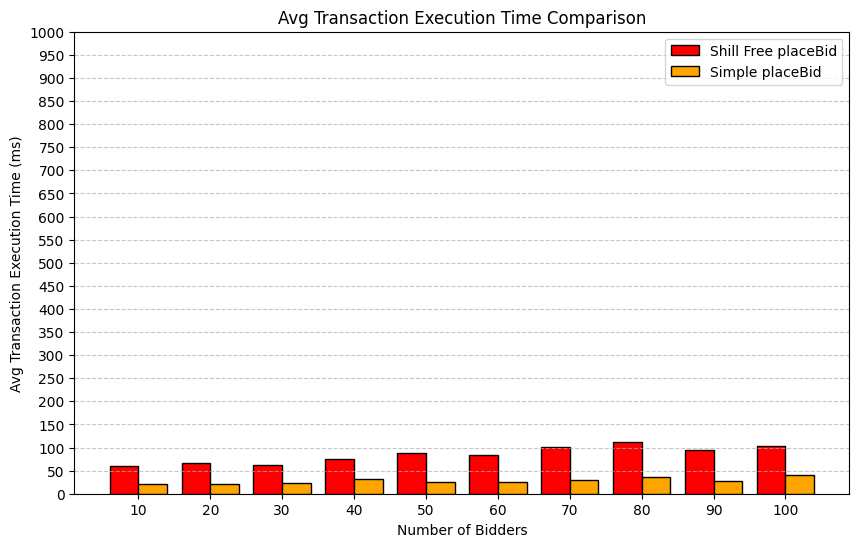}
        \caption{Scalability results}
    \label{fig:scalability}
    \end{subfigure}
    \caption{Evaluation of gas usage, latency, and scalability of system functions}
    
    \label{fig:cost_metrics}
\end{figure}

\section{Limitations}
While the proposed blockchain-based auction system effectively mitigates SB, certain limitations present opportunities for further research and development.

\subsection{Scalability Considerations}
The execution of operations on the Ethereum blockchain is faced with scalability challenges due to the cost of on-chain computation. As previously discussed in the performance evaluation, functions such as \texttt{placeBid} demonstrated increasing gas usage as bid volumes grew, reaching approximately 35,000 gas units for auctions with 100 bids. Despite optimisations using Solidity mappings and memory-efficient data structures, transaction fees remain a major challenge for large-scale deployment. To address this, future implementations can explore layer-2 solutions or alternative blockchains with lower transaction costs to improve scalability.

\subsection{Storage Overhead}
Storing full bidder histories and behavioural metrics on-chain results in substantial storage costs. As shown in our analysis, supporting 1,000 auctions with 100 bids each requires approximately 13.3~MB of on-chain data, making large-scale deployment economically challenging. While mappings optimise data access, the cost of maintaining large volumes of persistent data remains high.
A hybrid storage model could address this limitation by keeping only essential information—such as verification hashes—on-chain, while storing detailed bidding records off-chain. This approach can reduce on-chain storage requirements by over 90\% and significantly improve system scalability.

\subsection{Validation Opportunities}
Blockchain-based auctions are relatively new, and no real-world data exists to test SB patterns in decentralised environments. While the proposed metrics are derived from previous research on auction fraud in centralised environments, additional empirical testing would provide stronger validation for different auction scenarios. One possible future work would include developing simulation frameworks and collecting empirical data as blockchain auctions see more use.

\subsection{Deployment Variability}
Although initial results from simulation are promising, testing was conducted on a controlled local network. Real-world deployments on testnets or mainnets experience higher variability in gas prices, network congestion, and transaction confirmation times. Further testing in live environments is necessary to capture these dynamics and fine-tune system performance.

\subsection{Detection Refinement}
The transparency of blockchain transactions ensures trust but also allows shill bidders to observe and adapt their bidding strategies in order to avoid detection.
While the current model uses fixed-weight scoring for simplicity and verifiability, we acknowledge that such static configurations may be vulnerable to evolving adversarial behavior. To address this, future research will explore adaptive scoring mechanisms and feedback-based weight tuning to enhance resilience. Further extensions may also target more sophisticated forms of collusion beyond the nine core patterns addressed in this work, ensuring the system remains robust over time and effective in increasingly complex auction environments.

\subsection{User Experience Balance}
The penalty system applies to all participants, which may impact legitimate bidders who adopt aggressive  but honest bidding. While the current system provides a good balance between deterrence with usability, future improvements  can introduce new approaches that minimize false positives while maintaining strong fraud prevention capabilities.

These limitations do not detract from the contributions of this work but instead highlight potential directions for advancing secure and transparent auction systems. This framework represents a foundational step toward blockchain-based SB prevention, demonstrating its effectiveness in making fraudulent behaviour economically unfeasible.

\section{Conclusion and Future Work}
This paper presents a blockchain-based English auction system with an integrated anti-shill bidding mechanism. The proposed framework addresses one of the most challenging vulnerabilities in online auctions by using smart contracts to detect and deter SB in real time. The system makes three contributions to decentralized auction security.

First, we introduce a dynamic penalty mechanism based on the Bid Shill Score (BSS) and the dynamic penalty fee model, which evaluate nine distinct bidding patterns to identify and deter suspicious bidders. Unlike traditional post-auction analysis methods, this approach prevents manipulation in real time, making fraudulent bidding economically unviable. 
Simulation results confirmed the effectiveness of this approach, with median BSS values successfully distinguishing between honest and shill bidders across multiple advanced attack scenarios. Notably, coordinated multi-account and time-collusion attacks produced BSS scores exceeding 68--76\%, while honest bidders remained significantly lower.

Second, the implementation demonstrates that blockchain technology guarantees a transparent and tamper-proof auction ecosystem while maintaining decentralization. The consensus mechanism via smart contracts ensures applying the predefined rules, ensuring fairness for legitimate bidders. Testing confirms that the system effectively reduces the profitability of SB for sellers.

Third, the framework balances detection accuracy with computational efficiency, which is essential for real-world deployment. Performance evaluations showed that the bidding function maintained low latency, averaging around 300 ms even with 100 bids, while gas consumption remained within reasonable bounds. These results indicate that the approach is practical and scalable for typical auction volumes.

Future work will focus on further validation and optimization. We aim to explore an agent-based simulation environment with actual auction data to validate and refine the BSS accuracy.  Furthermore, we look into off-chain data storage methods and layer-2 scaling solutions to reduce gas costs while maintaining security guarantees, which will make the system more feasible for high-frequency auctions.
Furthermore, the model currently relies on fixed-weight scoring, which may be susceptible to adaptive adversaries. Future research will address this by investigating adaptive scoring and reinforcement-based tuning to enhance detection robustness. Additionally, full formal verification and live testnet deployments are planned to further validate the security and resilience of the proposed solution.

This study provides a foundation for secure and transparent decentralized auction systems. By making SB economically unfeasible, the proposed framework supports fair competition and enhances confidence in blockchain-based online commerce. As decentralized marketplaces continue to develop, this work contributes to ensuring their integrity and long-term sustainability.

\section{Acknowledgments}

We extend our sincere gratitude to Commed IA for their invaluable guidance and support throughout this project.

Mohamed Abdelhai Bouaicha is currently the beneficiary of a Ph.D. research grant partially funded by the Italian Ministerial Decree No. 352/2022 – PNRR, Mission 4, Component 2, "From Research to Enterprise" – Investment 3.3. "Introduction of innovative doctoral programs that respond to the needs of enterprises and promote the recruitment of researchers from enterprises" – an innovative doctoral scholarship with an industrial focus.

This research was also partially funded by the Apulian Italian Region within the project ``INVISIBLE BUSINESS - EXPONENTIAL TECHNOLOGIES FOR INVISIBLE BUSINESSES'' under the Regional Operational Programme ``P.O. PUGLIA FESR 2014–2020 – Asse I – Obiettivi specifici 1a Azione 1.1 Sub-Azione 1.1.b – Regolamento Regionale del 30 settembre 2014, n. 17 e s.m.i. – Titolo II – Capo 2 – Art. 26 – PROGRAMMI INTEGRATI DI AGEVOLAZIONE – PIA MEDIE IMPRESE'', funded through the ``Atto Dirigenziale'' n.686 – 20/09/2021.

\bibliographystyle{elsarticle-num}
\bibliography{ref}

\end{document}